\documentclass[aps,prl,floatfix,twocolumn,amsfonts,amssymb,amsmath,superscriptaddress]{revtex4-1}

\usepackage{float}
\usepackage{amsmath,amssymb,amsthm}
\usepackage{mathtools}
\usepackage{physics}
\usepackage{tikz}
\usepackage{graphicx}
\usepackage{microtype}

\renewcommand{\epsilon}{\varepsilon}

\newcommand{\vir}{\textsl{v}}
\newcommand{\sys}{\textsl{s}}
\newcommand{\anc}{\textsl{a}}
\newcommand{\rf}{\text{rot}}

\newcommand{\iu}{\mathrm{i}}
\newcommand{\eu}{\mathrm{e}}

\newcommand{\pol}{\epsilon}
\newcommand{\virpol}{\pol_\vir}

\newcommand{\gmax}{\gamma_{\max}}
\newcommand{\gmin}{\gamma_{\min}}
\newcommand{\ginf}{\gamma_{\inf}}
\newcommand{\gref}{\gamma_\rf}

\newcommand{\opel}[3]{{\bra{#1}}{#2}{\ket{#3}}}

\graphicspath{{./fig/}}

\begin{document}

\title{Cooling limits of coherent refrigerators}
\author{Rodolfo R. Soldati}
\affiliation{Institute for Theoretical Physics I, University of Stuttgart, D-70550 Stuttgart, Germany}
\affiliation{Instituto de Física da Universidade de São Paulo, 05314-970 São Paulo, Brazil}
\author{Durga B. R. Dasari}
\affiliation{3rd Institute of Physics, IQST, and Research Centre SCoPE, University of Stuttgart, Stuttgart, Germany}
\author{J\"{o}rg Wrachtrup}
\affiliation{3rd Institute of Physics, IQST, and Research Centre SCoPE, University of Stuttgart, Stuttgart, Germany}
\affiliation{Max Planck Institute for Solid State Research, Stuttgart, Germany}
\author{Eric Lutz}
\affiliation{Institute for Theoretical Physics I, University of Stuttgart, D-70550 Stuttgart, Germany}

\begin{abstract}
Refrigeration  limits are of fundamental and practical importance. We here show that quantum systems can be cooled below existing incoherent cooling bounds by employing coherent virtual qubits, even if the amount of coherence is incompletely known. Virtual subsystems, that do not necessarily correspond to  a natural eigensubspace of a system, are a key conceptual tool in quantum information science and quantum thermodynamics.  We derive universal coherent cooling limits and introduce specific protocols to reach them. As an illustration, we  propose a generalized algorithmic cooling protocol that outperforms its current incoherent counterpart. Our results provide a general framework to investigate the performance of coherent refrigeration processes. 
\end{abstract}

\maketitle

Partitioning a quantum system into subsystems with special features is an essential task in quantum theory. In many quantum information processing applications, it is, for instance, advantageous to encode quantum information in a subspace that is protected from the detrimental influence of the surrounding environment \cite{nie02}. Such a decomposition of the Hilbert space is associated with a given tensor product structure that is specified by the algebra of relevant observables \cite{zan01,zan04}. Oftentimes, this factorization does not correspond to the natural tensor product of spatially distinguishable systems, but rather of virtual ones \cite{lid13}. The general notion of virtual subsystems has found widespread use in quantum error correction \cite{kni99,kri05,kni06}, the theory of decoherence-free subspaces \cite{dua97,zan97,lid98} and noiseless systems \cite{zan03,choi06,blu08}, as well as in quantum computing \cite{dur03,cai09,cai10} and the study of entanglement \cite{pop05,kab20,car21}. Incoherent virtual qubits have, moreover, recently been successfully employed in quantum thermodynamics to investigate the properties of small quantum machines, such as heat engines and refrigerators, \mbox{and optimize their performance \cite{bru12,ven13,cor13,boh15,mit15,sil16,man17,erk17,du18,mit19,rig21}.} However, to our knowledge, the cooling advantage of coherent virtual qubits has not been examined so far.

Efficient cooling methods are crucial for the analysis of low-temperature quantum phenomena, from the physics of atoms and molecules \cite{met99,let09} to novel states of matter  \cite{mac92,ens05} and the design of quantum devices \cite{des09,hid21}. Low-temperature states with high purity are indeed a prerequisite for the implementation of quantum information processing algorithms that surpass their classical counterparts \cite{nie02}. Determining general refrigeration limits, independent of particular setups and cooling mechanisms, is hence of central importance \cite{ket92,wan13,wu13,tic14,all11,ree14,cli19,cli19a}. The steady-state temperature of a qubit with frequency $\omega$, valid for any cooling machine, is, for example, lower bounded by $T_\text{min}= T\omega/\Omega$, where $T$ is the bath temperature and $\Omega$ the largest energy of the machine \cite{all11,ree14,cli19,cli19a}. The temperature $T_\text{min}$ is believed to be the lowest possible qubit temperature achievable \cite{all11,ree14,cli19,cli19a}. Virtual subspaces have provided a unifying paradigm in this context, revealing that any refrigeration process may be understood as a generalized swap operation between the state of the system to be cooled and that of a sufficiently pure incoherent virtual subsystem of the environment \cite{tic14}.

We here demonstrate that coherent virtual qubits can be utilized as a resource to cool below existing incoherent refrigeration limits such as $T_\text{min}$ \cite{all11,ree14,cli19,cli19a}. Using a geometric representation of the cooling dynamics inside the Bloch sphere, we derive novel coherent cooling bounds that depend on the available  quantum coherence. We show how these general  bounds may be reached for  any standard purification mechanism. We observe that ground-state cooling is, in principle, possible in the limit of maximum coherence, provided perfect control over the system is given. We additionally establish that enhanced refrigeration is still achievable, when the amount of quantum coherence is incompletely known. We explicitly determine the corresponding cooling conditions and refrigeration limits. Furthermore, we apply our results to heat-bath algorithmic cooling, a powerful  cooling technique that employs standard quantum logic gates to extract heat out of a number of target spins in order to increase their polarization \cite{boy02,par16,fer04,sch05,sch07,rem07,kay07,bra14,rai15,rod16,rai19,sol22}. We concretely introduce a generalized cooling algorithm that involves a coherent virtual qubit, and show that it can outperform  incoherent algorithmic cooling protocols. Finally, we outline a general strategy to boost existing incoherent cooling methods.

\textit{Cooling with a coherent virtual qubit}. Let us consider a system qubit $s$ with Hamiltonian $H_\sys = \omega \sigma_z$, where $\omega$ is the frequency and $\sigma_z$ the standard Pauli operator. We denote by $\ket{0}$ ($\ket{1}$) its ground (excited) state. We assume that the system is   initially in a thermal state, $\rho_\sys(0) = \eu^{-\beta(\pol_\sys) H_\sys} / \tr[\eu^{-\beta(\pol_\sys) H_\sys}]$, where $\beta(\pol_\sys) = \omega^{-1}\ln[(1 + \pol_\sys)/(1 - \pol_\sys)]$ is the inverse temperature  and $\pol_\sys = \tr[\sigma_z\rho_\sys]$ the polarization. We describe a generic cooling process with a refrigeration superoperator, $\mathcal{R}_{t}[\rho_\sys(0)]$, that acts on the density operator of the system. Time can be either continuous or discrete. An example of a continuous-time process is that of a quantum refrigerator that cyclically runs between two heat reservoirs \cite{kos14}, in which case $\mathcal{R}_t[\rho_\sys(0)] = \exp(t \mathcal{L})\rho_\sys(0)$, with the Lindbladian $\mathcal{L}$ that describes the open dynamics of the system. An example of a discrete-time process is provided by heat-bath algorithmic cooling \cite{boy02,par16,fer04,sch05,sch07,rem07,kay07,bra14,rai15,rod16,rai19,sol22}, where heat is extracted from the system by repeatedly coupling it to ancilla qubits, collectively denoted by $a$, for which $ \mathcal{R}_{n}[\rho_\sys(0)] = \tr_\anc [\mathcal{K}_{n}( \rho_\sys(0) \otimes \rho_\anc(0) )]$, where $\mathcal{K}_{n}$ is the quantum channel associated with the cooling algorithm. In general terms, the action of the superoperator $\mathcal{R}_{t}$   decreases  temperature and entropy of the qubit, which leads to an increase of its polarization and purity towards one. 

\begin{figure}[t]
    \centering
    \includegraphics[width=0.95\columnwidth]{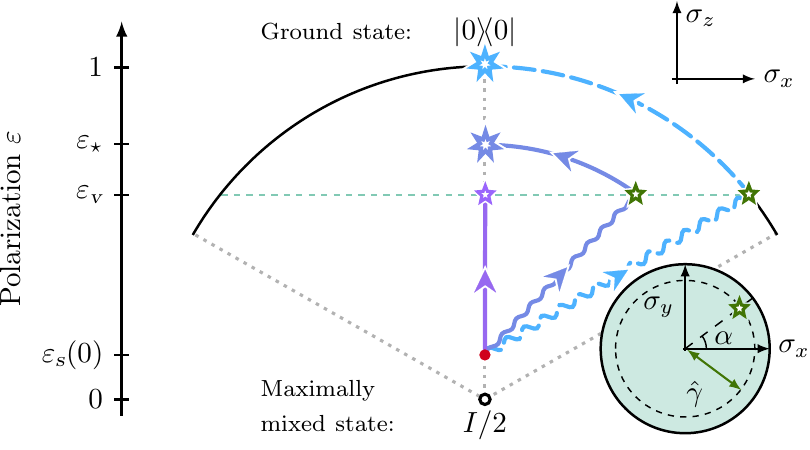}
    \caption{Bloch sphere representation of the refrigeration dynamics. Standard cooling processes may be regarded as a swap between the initial state of the system with polarization $\pol_\sys(0)$ (red dot) and an incoherent virtual qubit with polarization $\virpol$ (purple star). Adding coherence, characterized by the parameter $\gamma$, to the virtual qubit moves the state closer to the surface of the  sphere while keeping its polarization constant (green star). Better cooling is achieved by unitarily rotating the state back to the energy axis (violet star). Ground state cooling (blue star) is, in principle, possible for maximum coherence $(\gamma=1)$, when the state reaches the surface of the sphere. The inset shows the coordinates of the virtual qubit, radius $\hat{\gamma}
    = \gamma\sqrt{1 - \virpol^2}/2$ and angle $\alpha$, viewed from above.}
    \label{fig:swap-dynamics}
\end{figure}

From a physical point of view, for long times (or large cycle number), the cooling dynamics effectively exchanges the initial state $\rho_\sys(0)$ of the system qubit with the asymptotic state $ \mathcal{R}_{\infty}[\rho_\sys(0)]$. Mathematically, this operation may be regarded as a full swap \cite{nie02} between initial and final states. We shall use this observation to operationally define the virtual qubit. Virtual subsystems are often determined by specifying a   given factorization of the Hilbert space \cite{zan01,zan04,lid13,kni99,kri05,kni06,dua97,zan97,lid98,zan03,choi06,blu08,dua97,dur03,cai09,cai10,pop05,kab20,car21}, see also Ref.~\cite{com}. However, any purification dynamics may be viewed as a generalized swap process between the system qubit and a sufficiently pure virtual  qubit \cite{tic14}. By reversing this argument, we can identify the virtual qubit with the steady state    $ \rho_\vir=\mathcal{R}_{\infty}[\rho_\sys(0)]$. The virtual qubit is thus encoded in the refrigeration superoperator, hence providing an operational method to determine it for an arbitrary cooling process. This   is   nontrivial, since usually the virtual qubit  does not  coincide with any original qubit of the problem, but is given by two levels in a multidimensional system. For a continuous-time cooling process, the virtual qubit may correspond to a two-level subsystem of the two heat reservoirs considered as a composite entity \cite{bru12,ven13,cor13,boh15,mit15,sil16,man17,erk17,du18,mit19,rig21}, while it might be given by two levels  of the  ancilla qubit system for a discrete-time heat-bath algorithmic cooling process (see example below).
 Common (incoherent) cooling schemes are associated with incoherent virtual qubits.

We next analyze the cooling advantage of a coherent virtual qubit with linearly superposed energy states. To this end, it is convenient to describe the refrigeration dynamics using a Bloch sphere representation of the qubit, defined by writing any density operator as $\rho = (I + \sum_i i \sigma_i)/2$, with $i=(x,y,z)$ and $\sum_i i^2\leq 1$ \cite{ben06}. The process initially starts with the thermal state with coordinates $ (0,0,\pol_\sys(0)) $ (red dot in Fig.~1). Standard cooling protocols proceed along the (energy) $z$-axis \cite{kos14}, corresponding to an incoherent virtual qubit, $ \rho_\vir = (I + \virpol \sigma_z)/2 $, with polarization $ \virpol>\pol_\sys(0) $. After (infinitely) many cooling cycles, the system ends in the refrigerated state $(0,0, \virpol)$ (purple star in Fig.~1). The lowest possible temperature in this framework is $T_\text{min}= T\omega/\Omega$, where $T$ is the temperature of the (hot) bath and $\Omega$ is the largest energy of the machine \cite{all11,ree14,cli19,cli19a}. So far, strategies to improve this incoherent cooling limit amount to decrease the temperature of the virtual qubit, or, equivalently, increase its polarization. 

We here explore another method based on a coherent virtual qubit, which we parametrize as
\begin{equation}
    \rho_\vir =
    \frac{1}{2}
    \begin{pmatrix}
        1 + \virpol & \gamma \eu^{-\iu \alpha} \sqrt{1 - \virpol^2}   \\
        \gamma \eu^{ \iu \alpha} \sqrt{1 - \virpol^2}  & 1 - \virpol
    \end{pmatrix}, 
\end{equation}
  where $0 \leq \gamma \leq 1$ is a measure of quantum coherence, and $\alpha$ defines the direction of $\sigma_\alpha = \cos(\alpha) \sigma_x + \sin(\alpha) \sigma_y$. The plane $\sigma_z \cross \sigma_\alpha$ is then perpendicular to the $\sigma_x \cross \sigma_y$ plane. The Bloch coordinates of the virtual qubit are now $(x_\vir + \iu y_\vir = \gamma \eu^{\iu \alpha}\sqrt{1 - \virpol^2},z_\vir=\virpol)$ (green star in Fig.~1). Endowing the virtual qubit with coherence $\gamma$ (with fixed polarization $\virpol$)  increases its purity, moving the virtual state away from the energy axis and closer to the surface of the Bloch sphere. After many cooling cycles, the system state will approach the coherent virtual state (wavy lines). Lower system qubit temperatures  may then be achieved by  rotating the system state back to the energy basis at the end of the coherent cooling process (violet star in Fig.~1). For this single additional step, we only consider unitary operations that do not change the purity. This is justified because any other resource that may further increase purity could have been used to achieve a higher  polarization $\virpol$ in the first place \cite{com1}.

\textit{Coherent cooling limits.} The maximum achievable system polarization using a coherent virtual qubit is 
\begin{equation}
\label{2}
\pol_\star(\gamma) = \sqrt{\virpol^2 + (1 - \virpol^2) \gamma^2},
\end{equation}
given by the eigenvalues of the coherent virtual qubit state (violet star in Fig.~1). This coherent cooling limit neither depends on the details of the cooling protocol  nor on the specific mechanism creating the coherent virtual qubit; it hence appears to be universal. The coherent bound $\pol_\star$ generally exceeds the incoherent bound $\virpol$. The unitary transformation realizing the appropriate rotation at the end of the coherent cooling sequence should be orthogonal to $\sigma_\alpha$, with an angle that depends on $\gamma$. It is explicitly given by $V = \exp(\iu \chi(\gamma)\, \sigma_{\perp} ( \alpha)/2)$ with $\sigma_{\perp} (\alpha) = -\sin(\alpha) \sigma_x + \cos(\alpha) \sigma_y $ and $\chi(\gamma) = \arccos[\virpol/\pol_\star(\gamma)]$ (violet line in Fig.~1). The standard refrigeration bound, $\pol_\star(0) = \virpol$, is recovered in the absence of coherence, $\gamma = 0$. In the opposite limit of maximum coherence, $\gamma = 1$, ground state cooling, $\pol_\star(1) = 1$, is in principle possible, with a rotation angle $\chi(1) = \arccos(\virpol)$ (blue line in Fig.~1). We emphasize that this result does not contradict the unattainability principle of the third law of thermodynamics \cite{bel04}, since it presupposes perfect knowledge of the values of $\gamma$ and $\alpha$, as well as  a perfect implementation of the unitary rotation. Both are not possible in any realistic experiment. This is reminiscent of the (practical) impossibility of violating the second law  of thermodynamics by reversing all the velocities of a system (Loschmidt's reversibility paradox) \cite{bel04}, since this would require perfect control over the system.
\begin{figure}
    \centering
    \includegraphics[width=0.95\columnwidth]{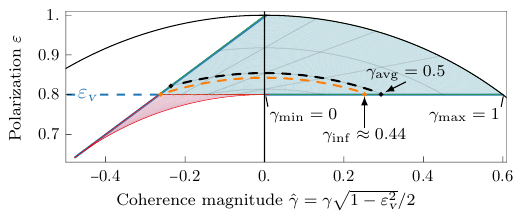}
    \caption{Improved cooling with incomplete knowledge of the amount of quantum coherence of the virtual qubit. Enhanced cooling, $\pol\geq \virpol$, can be achieved even if the amount of coherence $\gamma$ is only known to be in an interval $[\gmin, \gmax]$, when the virtual qubit state is such that $\ginf<\gamma$, Eq.~\eqref{5} (blue shaded area). This even holds when the value of $\gamma$ is completely unknown ($\gmin=0$ and $\gmax=1$). This is, in particular, the case, for the middle point, $\gamma_\text{avg}$, of the confidence interval. }
    \label{fig:coh-advantage}
\end{figure}

In practice, the parameters $\gamma$ and $\alpha$ are usually not known precisely. Yet, improved cooling might still be achieved, as we will now discuss \cite{com4}. Let us distinguish between the parameters $(\gamma, \alpha)$ of the given coherent virtual state and the parameters $(\gref, \alpha_\rf)$ of the implemented rotation $V$. We  assume, for simplicity, that $\alpha = \alpha_\rf$ is fixed and that $\gamma$ is uniformly distributed in a  confidence interval $[\gmin, \gmax]$ (the case of an unknown $\alpha$ is examined in the Supplemental Material \cite{sm}). We first determine the maximal value $\gref$ that still yields improved cooling given an amount of coherence $\gamma$. There are three regions for which this is possible. For $1/2 \leq \gamma$, every choice of $\gref$ leads to improved cooling, irrespective of the value of $\virpol$. For $\gamma \leq 1/2$, increased refrigeration will happen for all values of $\gref$, provided that the polarization obeys $\virpol \leq \gamma / (1 - \gamma)$. Finally, for $\virpol > \gamma / (1 - \gamma)$, enhanced cooling \mbox{will occur for}
\begin{equation}
   \gamma_{\rf}\leq \frac{
        2 \virpol^2 \gamma
 }{
        \virpol^2 - \gamma^2 (1 - \virpol^2)
    } .
\end{equation}
The maximum achievable system polarization depends on the  rotation angle $\chi(\gref) > 0$, and  is given by
\begin{equation}
\label{4}
    \pol_{\chi(\gref)} = \frac{
        \virpol^2 + \gamma \gref (1 - \virpol^2)
    }{
        \sqrt{\gref^2 + \virpol^2 (1 - \gref^2)}
    } .
\end{equation}
When $\gref \leq \gmin$, enhanced cooling is possible for all $\gamma > \gmin$. The highest possible system polarization, $\pol_{\chi(\gmax)}$, is obtained for $\gamma = \gmax$. Care is required when $\gref>\gmin$, since a large enough rotation of a state with small enough coherence $\gamma$ may result in heating (red shaded area in Fig.~2). Equating $\pol_{\chi(\gref)}$, Eq.~\eqref{4}, to $\virpol$, and solving for $\gamma$ yields a second lower bound for the coherence $\gamma$, that is different from $\gmin$,
\begin{equation}
\label{5}
    \ginf(\gref) =
    \virpol
    \frac{
        \sqrt{
                \virpol^2 + \gref^2 (1 - \virpol^2 )
            }
        }{
            \gref (1 - \virpol^2)
        }
    - \frac{\virpol^2}{\gref (1 - \virpol^2)}
    ,
\end{equation}
so that heating is avoided (Fig.~2). Best refrigeration is  found for $\gamma = \gmax$ and is again given by Eq.~\eqref{4} \cite{com2}. We note that, for $\gref \leq \gmax$, taking the unknown value of $\gamma$ as the middle point of the confidence interval,  $\gamma_\text{avg} = (\gmin + \gmax)/2$, always leads to improved cooling, since  $\gamma_\text{avg}> \ginf$ (Fig.~2). Interestingly, this result even holds when the amount of coherence $\gamma$ is completely unknown ($\gmin=0$ and $\gmax=1$). The above equations fully determine the improved quantum cooling domain 
 (blue area in Fig.~2). They establish the robustness of the proposed scheme against experimental imperfections.

\textit{Application to heat-bath algorithmic cooling.} The above analysis of the cooling advantage of coherent virtual qubits is completely general. We next apply it to  heat-bath algorithmic cooling \cite{boy02,par16,fer04,sch05,sch07,rem07,kay07,bra14,rai15,rod16,rai19,sol22} that has recently been implemented experimentally \cite{bau05,rya08,par15,ata16,zai21}. Its smallest version is made of three qubits: one system (target) qubit with state $\rho_\sys(n) = \rho_1(n)$, to be cooled, and two ancilla (reset) qubits with separable states $\rho_\anc = \rho_2 \otimes \rho_3$. The cooling algorithm consists of (i) a compression step during which a unitary $U$ implements the transition $\ketbra{011}{100} + \text{h.c.}$, pumping heat  out of the target qubit and into the two reset qubits, followed by (ii) a refresh step that thermalizes the reset qubits back to their initial bath polarization $\pol_2$ and $\pol_3$ (Fig.~3a). The corresponding quantum channels, that operate on the three-qubit ensemble, are  $\mathcal{U}(\rho) = U \rho U^\dag$ for the cooling part and $\mathcal{E}(\rho) = \rho \otimes \rho_2 \otimes \rho_3$ for the expansion part. In the limit of many cooling cycles, the target qubit reaches the asymptotic polarization $\pol_1(\infty)$. This operation may be understood as a full swap between the initial state of the target qubit and an incoherent virtual qubit consisting of the two states $\ket{0_2 0_3}$ and $\ket{1_2 1_3}$. The standard cooling limit is here determined by the polarization of the incoherent virtual qubit, $\virpol(\pol_2, \pol_3) = (\pol_2 + \pol_3)/( 1 + \pol_2 \pol_3)$ \cite{sol22}. For simplicity, we will set $\pol_2 = \pol_3 = \pol_\anc$ in the remainder.

 We now introduce a coherent extension of this  cooling algorithm such that $\mathcal{E}^c(\rho) = \rho \otimes \rho_{23}$, where $\rho_{23}$ is a  nonseparable state. Such state may, for example, be simply generated  by considering interacting reset qubits (see detailed discussion below), or by directly entangling them \cite{har22}. The dynamics of the system qubit is then described by the coherent refrigeration process
\begin{equation}
\label{6}
    \rho_1(n) = \mathcal{R}^c_n(\rho_1(0)) = ({\tr_{23}} \circ \mathcal{U} \circ \mathcal{E}^c)^n(\rho_1(0)) ,
\end{equation}
after $n$  cycles. The transformation \eqref{6} both cools the target qubit and transfers coherence from the virtual qubit to the target qubit. In the asymptotic limit, the coherent refrigeration superoperator takes the form of a swap with a coherent virtual qubit, $\mathcal{R}^c_\infty(\cdot) = \tr_{\vir} S (\cdot \otimes \rho_\vir) S^\dag$, with swap operator $S$ \cite{nie02}. We parametrize the coherence of the virtual qubit  by the matrix elements, $\mel{00}{\rho_{23}}{11} = ({\xi}/{4}) \eu^{-\iu \alpha} ( 1 - \pol_\anc^2 ) $, ($0 \leq \xi \leq 1$), which  implies $\mel{0}{\rho_\vir}{1} = ({\gamma}/{2}) \eu^{-\iu \alpha} \sqrt{1 - \virpol^2}$, with $\gamma = \virpol \xi$. In order  to achieve a fair comparison between coherent and incoherent cooling strategies, we analyze the effect of coherence while fixing the polarization of the virtual  qubit so that it is the same as that of the incoherent one.  In view of Eq.~\eqref{2}, the maximum  achievable target polarization of this coherent  algorithmic  cooling scheme is then $ \pol_\star = \virpol \sqrt{ 2 - \virpol^2 } \geq \virpol$, \mbox{for $\xi=1$}, which clearly exceeds the current incoherent cooling limit $\virpol$.

\begin{figure}[t]
    \centering
    \includegraphics[height=0.6\columnwidth]{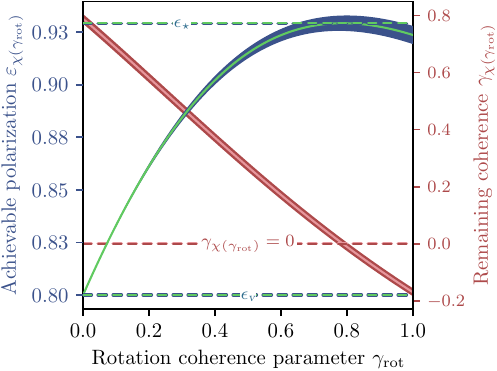}
    \caption{Achievable target polarization $ \pol_{\chi(\gref)} $, Eq.~\eqref{4}, (green) after a  unitary rotation parametrized by the coherence parameter $\gref$  for a realistic heat-bath algorithmic cooling protocol using nitrogen-vacancy (NV) centers in diamond. The coherent virtual qubit has polarization $\virpol  =0.8$. Maximum polarization $\pol_\star$ is reached for $\gref\sim 0.79$, a value for which the remaining coherence of the virtual qubit, $ \gamma_{\chi(\gref)}$, (pink) vanishes; the blue and red shaded areas represent the respective confidence intervals of $ \pol_{\chi(\gref)} $ and $ \gamma_{\chi(\gref)}$. }
    \label{fig:hbac-coh-interval}
\end{figure}

As an illustration, we examine the achievable polarization $ \pol_{\chi(\gref)} $, Eq.~\eqref{4}, as a function of $\gref$ for a realistic confidence interval available in experiments using  nitrogen-vacancy  (NV) centers in diamond \cite{zai21}. The  two reset qubits are here  two carbon $^{13}$C nuclear spins that are coupled to the central electron spin of the NV center \cite{zai21}.  We consider a maximally coherent virtual qubit ($\xi = 1$) with polarization $\virpol = \gamma =0.8$ prepared with $97.6\%$ fidelity. We accordingly have $\gmin = 0.78$ and $\gmax = 0.8$. Figure~3 displays the tradeoff between the achievable polarization $ \pol_{\chi(\gref)}$ (green) and the remaining coherence $ \gamma_{\chi(\gref)}$ (pink) after the final unitary rotation (the blue and red shaded areas represent the corresponding confidence intervals). Maximal polarization of  $ \pol_{\chi(\gref)} \sim 0.93$ is attained for $\gref \sim 0.79$, where the remaining coherence vanishes. This value significantly exceeds the incoherent cooling bound $\virpol=0.8$ and saturates the coherent refrigeration bound $\pol_\star$. For a rotation angle larger than the optimal value, the achievable polarization decreases; this overshoot corresponds to the red shaded area on the left-hand side of Fig.~2.

\textit{Roadmap to enhanced quantum cooling.} The above results outline a generic strategy to improve  existing incoherent cooling schemes which consists of three steps: (i) determining the usually nontrivial incoherent virtual qubit from the steady-state limit of the incoherent refrigerator superoperator, (ii)  endowing it with quantum coherence, which can be done passively by choosing an appropriate set of controlled observables, and (iii) implementing an appropriate unitary rotation on the system qubit at the end of the cooling sequence to bring it back to the (incoherent) energy axis. The first step  holds for any cooling protocol, and allows one to identify the incoherent virtual qubit from the knowledge of the incoherent cooling process. The second step will in general depend on the considered cooling method. However, the heat-bath algorithmic cooling example analyzed above provides valuable general clues. In the latter case, the coherent virtual qubit may indeed be easily generated by considering interacting reset qubits, for instance, via a transverse-Ising type coupling of the form $H_{23}= -(\omega/4) (\sigma^z_2 + \sigma^z_3) + g \sigma^x_2 \sigma^x_3 - g \sigma^y_2 \sigma^y_3$ with coupling constant $g$ (the corresponding coherent cooling process is simulated in the Supplemental Material \cite{sm}); the standard heat-bath algorithmic cooling protocol uses noninteracting reset qubits with $g=0$ \cite{bau05,rya08,par15,ata16,zai21}. The system will then exhibit quantum coherence in the local energy bases of each reset qubit when  the global two-qubit state is an (incoherent)  thermal Gibbs state, which, in turn, will make the virtual qubit coherent. This follows from the fact that quantum coherence generally depends on the choice of a basis in Hilbert space \cite{nie02}, which  is reminiscent of the finding that the tensor-product structure of the Hilbert space generally depends on the choice of a set of observables \cite{zan04}. The same mechanism should be applicable to more complicated refrigeration schemes; it is important to realize that all the other parts of the cooling algorithm are not modified. Finally, the last step only entails a single unitary rotation of the system qubit after completion of the coherent refrigeration process.

\textit{Conclusions.}  The current experimental development of quantum applications seems to mostly rely on nonquantum cooling schemes, without fully harnessing the power of coherent refrigeration.  By extending the concept of virtual qubits to coherent virtual qubits, and further showing that they may be operationally determined from the asymptotic limit of any refrigeration superoperator, we have shown that it is possible to cool beyond existing incoherent refrigeration bounds. We have concretely obtained generic coherent cooling bounds that are independent of the considered setup. Quantum coherence thus appears as a useful physical resource in this context. Remarkably, enhanced cooling is feasible, even if the amount of coherence is not exactly known. This makes this quantum refrigeration scheme robust for experimental implementations.  In the present approach, virtual qubits function as a reservoir of coherence, which is different from  dynamically generated coherence  that has been found to assist  cooling in some instances, such as single-shot cooling \cite{mit15}. We have moreover introduced a coherent heat-bath algorithmic cooling protocol and  specified its enhanced cooling limit. 
Our findings provide a general framework to investigate the properties of coherent refrigeration processes and additionally improve the performance of existing incoherent cooling protocols.

 \textit{Acknowledgments.} We acknowledge the financial support by the DFG
 (FOR 2724), DFG (509457256), BMBF (Grant No. 16KIS1590K), EU GRK2642, AMADEUS,
 QIA, Max Planck Society, and the Baden-W\"urttemberg Foundation. R. R. S.
 further acknowledges fruitful discussions with R. Laflamme and N. A. Rodriguez-Briones,
 and the financial support by  DAAD 
 (research grant Bi-nationally Supervised Doctoral Degrees/Cotutelle, 57507869) and from CNPq (141797/2019-3).

\clearpage
\onecolumngrid
\appendix

\renewcommand{\figurename}{Figure}
\renewcommand{\theequation}{S\arabic{equation}}
\renewcommand{\thefigure}{S\arabic{figure}}
\renewcommand{\bibnumfmt}[1]{[S#1]}

\newcommand{\opn}[1]{\operatorname{#1}}

\begin{center}
    \textbf{\large Supplemental Material}
\end{center}
\setcounter{equation}{0}
\setcounter{figure}{0}
\setcounter{table}{0}
\setcounter{page}{1}

The Supplemental Material contains details about (I) the dynamics of the target
qubit for the minimal heat-bath algorithmic cooling model with a coherent
virtual qubit, (II) the generation of the coherent virtual qubit, as well as
(III) the simulation of the generalized protocol, (IV) the determination of the
cooling regions for the case of an unknown coherence phase $\alpha$, (V) an
analysis of the thermodynamic performance of the coherent protocol, and (VI) a
comparison with  heat-bath algorithmic cooling with multiple reset qubits.

\section{Analytical solution of the target qubit dynamics in the natural representation}

We here provide the analytical solution of the target qubit dynamics for the
minimal heat-bath algorithmic cooling model with a coherent virtual qubit
discussed in the main text. To this end, we first introduce the natural
representation (vectorization map) \cite{wat18}, also sometimes referred to as
Liouville space representation \cite{gya20},  used to reach the result.

\subsection{Vectorization method}

We introduce the vectorization transformations that map states into column
matrices and quantum channels into matrix operators, acting on the columns. To
set the notation, we define vectorization on states as
\begin{equation}
    \opn{vec} \colon \rho \mapsto \vec{\rho} .
\end{equation}
In the qubit case, the central Hilbert space from which the quantum dynamics are
constructed is $\mathbf{H} = \mathbb{C}^2$. Thus, the density matrix of a state
is represented by an element of $\mathbb{M}_\mathbb{C}^{2\times 2}$, that is, by 2-by-2
matrices with complex entries that are Hermitian and positive
semidefinite. In terms of its elements, a density matrix is mapped by
$\opn{vec}$ by stacking its columns into a single one, such as in
\begin{equation}
    \frac{1}{2}
    \begin{pmatrix}
        1 + \epsilon & x + \iu y \\
        x - \iu y & 1 - \epsilon
    \end{pmatrix}
    \xmapsto{\opn{vec}}
    \frac{1}{2}
    \begin{pmatrix}
        1 + \epsilon \\ x - \iu y \\
        x + \iu y \\ 1 - \epsilon
    \end{pmatrix} ,
\end{equation}
where $\mathbb{M}_\mathbb{C}^{2\times 2} \stackrel{\opn{vec}}{\cong}
\mathbb{M}_\mathbb{C}^{1\times 4} = \mathbb{C}^4$. We call Liouville space the
resulting column matrix space.

The second vectorization mapping, also known as the natural representation of
quantum maps $\Phi$ \cite{wat18}, is induced from the following compatibility
condition:
\begin{equation} \label{eq:nat-rep}
    \Phi_\mathcal{K} \vec{\rho} = \opn{vec}[ \mathcal{K}(\rho) ] ,
\end{equation}
where $\mathcal{K}$ is a quantum channel, such as the one defined by the
operator sum representation with Kraus operators $K_\mu$ as $ \mathcal{K}(\rho)
= \sum_\mu K_\mu \rho K_\mu^\dag$. In words, the compatibility condition is the
statement that there exists $\Phi_\mathcal{K}$ such that it maps the vectorized
input state to the vectorization of the output of $\mathcal{K}$.

The solution to the compatibility condition in these variables is
\begin{equation}
    \Phi_\mathcal{K} = \sum_\mu K_\mu \otimes K_\mu^* .
\end{equation}
As such, a map $\Phi_\mathcal{K}$ is a matrix element of
$\mathbb{M}_\mathbb{C}^{4\times 4}$ that allows rewriting the operator sum
representation of $\mathcal{K}$ as simple matrix multiplication.

\subsection{Solution for target qubit evolution}

\begin{figure}[t]
    \centering
    \includegraphics{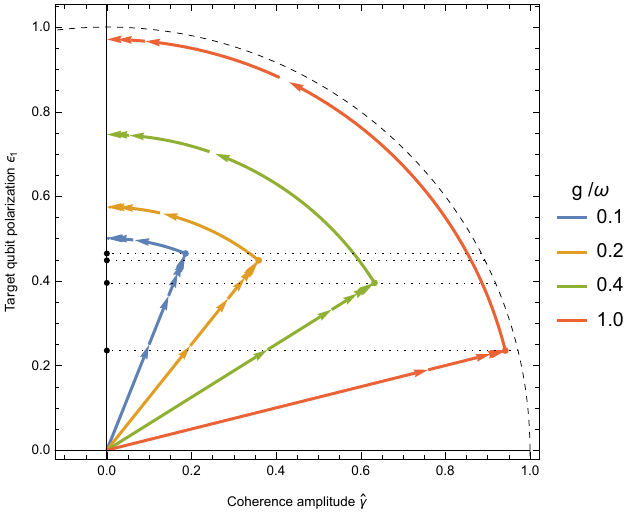}
    \caption{Bloch sphere representation of the evolution of the polarization
    $\epsilon_1$ of the target qubit as the number of cycle number $n$
    (arrowheads) is increased, for various values of the ratio $g/\omega$. The
    reset qubits are here assumed to interact via a transverse-field Ising
    Hamiltonian of the form $H = -\frac{\omega}{4} (\sigma^z_2 + \sigma^z_3) + g
    \sigma^x_2 \sigma^x_3 - g \sigma^y_2 \sigma^y_3$,  with frequency $\omega$
    and coupling constant $g$. Both the polarization and the amount of coherence
    increase with the number of cycles. The inverse temperature is taken to be
    $\beta = 1.09/\omega$. Black dots represent the polarization of the
    corresponding incoherent virtual qubit we compare to.}
    \label{fig:ising}
\end{figure}

The dynamical map $\mathcal{R}_n$ for heat-bath algorithmic cooling is
Markovian, and thus at arbitrary cycle number $n$, it contains every necessary
information of the qubit evolution. Its nonzero elements are \cite{sol22}
\begin{align}
    (\Phi_\mathcal{R}^n)_{11} ={}&
    \frac{
        (1 - \pol_2) (1 - \pol_3)
    }{
        2 (1 + \pol_2 \pol_3)
    }
    \frac{
        (1 - \pol_2 \pol_3)^n
    }{
        2^n
    }
    + \frac{
        (1 + \pol_2) (1 + \pol_3)
    }{
        2 (1 + \pol_2 \pol_3)
    } \\
    (\Phi_\mathcal{R}^n)_{21} ={}&
    \bigg[
        (\pol_2 + \pol_3) (\pol_2\pol_3 - 1)
    \notag \\
    &- \big[
            (\pol_2 + \pol_3) (1 - \pol_2\pol_3)
            - n (1 - \pol_2) (1 - \pol_3) (1 + \pol_2\pol_3)
        \big]
        \frac{
            (1 - \pol_2 \pol_3)^n
        }{
            2^n
        }
    \bigg] %
    \frac{
        \eu^{\iu \alpha'} \xi \sqrt{1 - \pol_2^2} \sqrt{1 - \pol_3^2}
    }{
        2 (\pol_2\pol_3 - 1) (1 + \pol_2\pol_3)^2
    } .
\end{align}
These are two elements that will help define the virtual qubit $\rho_\vir$
asymptotically as $n \to \infty$. As such, the rest of the matrix is
\begin{equation}
    \Phi_\mathcal{R}^n =
    \begin{pmatrix}
        (\Phi_\mathcal{R}^n)_{11} & 0 & 0 & (\Phi_\mathcal{R}^n)_{11} \\
        (\Phi_\mathcal{R}^n)_{21} & (\Phi_\mathcal{R}^n)_{22} & 0 & (\Phi_\mathcal{R}^n)_{21} \\
        (\Phi_\mathcal{R}^n)_{21}^* & 0 & (\Phi_\mathcal{R}^n)_{33} & (\Phi_\mathcal{R}^n)_{21}^* \\
        1 - (\Phi_\mathcal{R}^n)_{11} & 0 & 0 & 1 - (\Phi_\mathcal{R}^n)_{11}
    \end{pmatrix} .
\end{equation}
The remaining elements $(\Phi_\mathcal{R}^n)_{22}$ and
$(\Phi_\mathcal{R}^n)_{33}$ are explicitly given by
\begin{equation}
    (\Phi_\mathcal{R}^n)_{22} = (\Phi_\mathcal{R}^n)_{33} =
    \frac{ (1 - \pol_2\pol_3)^n }{ 2^n } .
\end{equation}
These elements carry information of initial coherences but vanish
asymptotically. However, in the transient dynamics, there are cross
contributions from the coherence that was input into the target and the
coherence that is yet to come. The full evolution of the target qubit is
accordingly
\begin{align}
    \opel{0}{\rho_1(n)}{0} &=
    \frac{
        (\pol_1 \pol_2 \pol_3 + \pol_1 - \pol_2 - \pol_3)
    }{
        2 (1 + \pol_2 \pol_3)
    }
    \frac{
        (1 - \pol_2 \pol_3)^n
    }{
        2^n
    }
    + \frac{
        (1 + \pol_2) (1 + \pol_3)
    }{
        2 (1 + \pol_2 \pol_3)
    } \\
    \opel{1}{\rho_1(n)}{0} &=
    \begin{multlined}[t]
        \left[
            \frac{
                (\pol_2 + \pol_3)
            }{
                (1 + \pol_2\pol_3)
            }
            - \left(
                \frac{
                    (\pol_2 + \pol_3)
                }{
                    (1 + \pol_2\pol_3)
                }
                - \frac{n}{1 - \pol_2 \pol_3}
                (\pol_1 \pol_2 \pol_3 + \pol_1 - \pol_2 - \pol_3)
            \right)
            \frac{
                (1 - \pol_2 \pol_3)^n
            }{
                2^n
            }
        \right] %
        \frac{
            \eu^{\iu \alpha'} \xi \sqrt{1 - \pol_2^2} \sqrt{1 - \pol_3^2}
        }{
            2 (1 + \pol_2\pol_3)
        } \\
        + \frac{\chi \sqrt{1 - \pol_1^2} }{2}
        \frac{
            (1 - \pol_2 \pol_3)^n
        }{
            2^n
        } ,
    \end{multlined}
\end{align}
where we here consider the presence of an initially coherent target qubit with
coherence $\chi = |\chi| \eu^{-\iu \arg(\chi)}$ parametrized as usual in the
rest of the paper. This contribution comes from $(\Phi_\mathcal{R}^n)_{22}$, and
vanishes asymptotically as it is replaced by the virtual qubit coherence.

\section{Generation of a coherent virtual qubit}

In the case of the three-qubit heat-bath algorithmic cooling example
discussed in the main text, a coherent virtual qubit  could be simply and
naturally created by considering two interacting reset qubits, instead of the
two noninteracting reset qubits employed in the standard, incoherent heat-bath
algorithmic cooling protocol  (coherence could, of course, also be engineered
artificially, but this would be more costly). For instance, the thermal state of
a two-spin transverse-field Ising model with Hamiltonian, $H = -\frac{\omega}{4}
(\sigma^z_2 + \sigma^z_3) + g \sigma^x_2 \sigma^x_3 - g \sigma^y_2 \sigma^y_3$,
naturally contains coherences with respect to  each qubit's local energy bases.
As a result, the corresponding virtual qubit also exhibits coherences that
scale  with  inverse temperature $\beta$  and coupling strength $g$ as $\gamma
\sim g \tanh{( \beta \sqrt{g^2 + \omega^2} )}$.

Figure~\ref{fig:ising} shows the Bloch sphere representation of the
evolution of the polarization $\epsilon_1$ of the target qubit  as the number of
cycle number $n$ (dots) is increased, for various values of the ratio
$g/\omega$. This figure is a concrete illustration (for each coupling
strength $g$) of the schematic diagram shown in Fig.~1 of the main text,
where we compare the asymptotic  state  with its corresponding
incoherent virtual qubit state, lying on the $y$-axis. We note that both the
polarization and the amount of coherence increase with the number of cycles,
indicating that the protocol  is capable of transporting quantum coherence from
each virtual qubit to the target qubit at the same time that the target qubit is
cooled. The surface of the Bloch sphere, corresponding to maximal target
coherence, is almost reached for $g/\omega=1$.

\section{Simulation of the coherent cooling protocol}

\begin{figure}[t]
    \centering
    \includegraphics{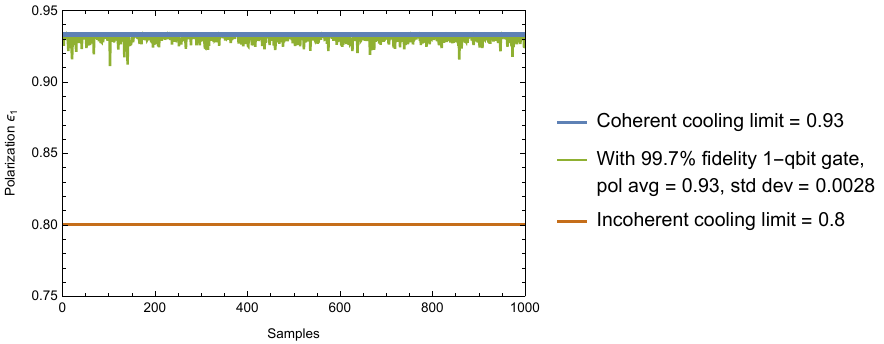}
    \caption{Target polarization for coherent and incoherent cooling
    protocols with realistic experimental parameters. Reset polarizations are
    set to $\epsilon_\textrm{res} = 0.5$ corresponding to an incoherent cooling
    limit of $\epsilon^\text{inc}_\textrm{lim} = 0.8$ (brown line). (1) Green:
    target polarization achieved with a 1-qubit unitary gate with fidelity
    $99.7\%$ applied once to the target qubit; it is close to the maximal
    achievable coherent polarization bound $\epsilon^\text{coh}_\textrm{max} =
    0.93$ (blue line).}
    \label{fig:error-estimate}
\end{figure}

In this section, we present a numerical simulation of the generalized
 heat-bath algorithmic cooling scheme with a coherent virtual qubit, using
 realistic experimental parameters. The nontrivial advantage of the suggested
 protocol is  that it is capable of transporting the coherences from each
 virtual qubit to the target qubit at the same time that the target qubit is
 cooled. This results in a target qubit which is also coherent in the steady
 state, and therefore a single unitary gate on the target qubit is enough to
 take advantage of all available coherences -- and achieve better cooling.

\begin{figure}
    \centering
    \includegraphics{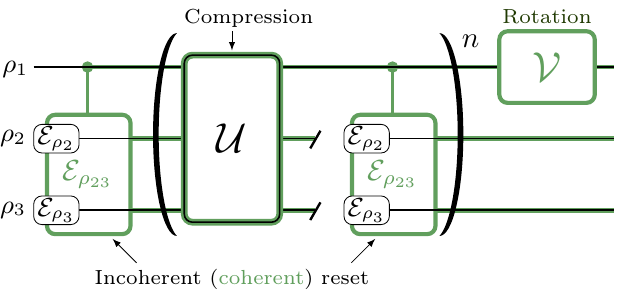}
    \caption{Minimal heat-bath algorithmic cooling application. Wire diagram for an $n$-iteration of the incoherent algorithm (black) and the coherent extension (green), followed by the polarization-enhancing rotation, $\mathcal{V}(\cdot) = V \cdot V^\dag$. The fidelity of the two-qubit gate that comprises part of the compression unitary U, as well as the particular form of this unitary, determine the coherence confidence interval discussed in the main text.}
    \label{fig:circuit}
\end{figure}

Figure~\ref{fig:error-estimate} shows the outcome of numerical  simulations
based on our available NV center gate fidelities for 1-qubit gates $\mathcal{V}$
(99.6$\%$ to 99.8$\%$); we assume 100$\%$ fidelity for the implementation of the
refrigeration channel $\mathcal{R}$ in order to better highlight the effect the
different gate fidelities. We observe that our protocol (green, see also
Fig.~\ref{fig:circuit}) gets very close to the coherent cooling limit
$\epsilon^\text{coh}_\textrm{max} = 0.93$ (blue), which clearly surpasses the
incoherent cooling limit $\epsilon^\text{inc}_\textrm{lim} = 0.8$ (brown).

\section{Cooling with unknown coherence phase}

\begin{figure*}[t]
    \centering
    \includegraphics{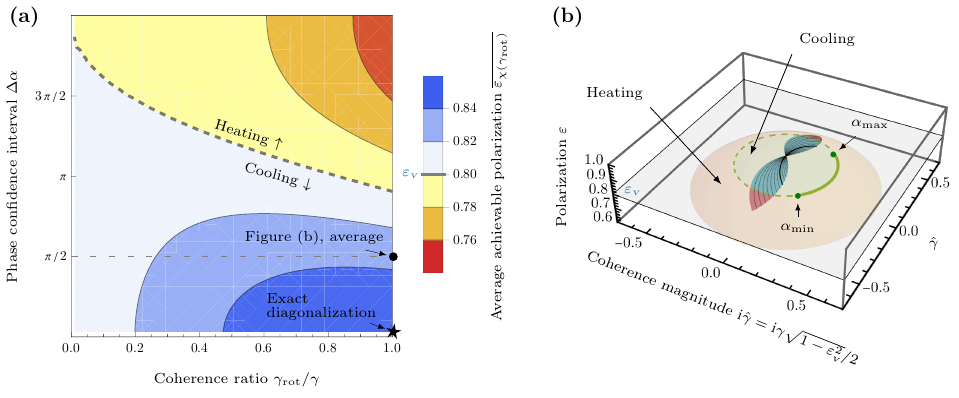}
    \caption{Achievable target polarizations for unknown coherence phase. (a) Average
    achievable polarization $\overline{\pol_{\chi(\gref)}}$ of the target qubit over ensemble of
    states, Eq.~\eqref{12}, for $\gref / \gamma < 1$ and various confidence
    intervals $\Delta\alpha = \alpha_{\max} - \alpha_{\min}$, with $\alpha_\rf=
    \alpha -\pi/2$ and $\virpol=0.8$. Average cooling is found for all values $\gref /
    \gamma$ when $\Delta \alpha < 0.9 \pi$. (b) Geometric representation (full
    green line) of ensemble of states for $\gamma = 0.5$ with a confidence
    interval of $\alpha_{\min} = -\pi/4$ and $\alpha_{\max} = \pi/4$. The shaded
    region of achievable cooling (blue) and heating (red) is found by applying
    the unitary rotation with parameters $\gref = \gamma$ and $\alpha_\rf = -\pi/2 +
    \alpha$ for every point in the confidence interval (see corresponding
    average in (a)). The thicker black line represents the achievable target 
    polarization region for fixed phase $\alpha_\rf = (\alpha_{\max} - \alpha_{\min})
    / 2$.}
    \label{fig:alpha-interval}
\end{figure*}

We next consider the situation where the coherence angle $\alpha$ is known to be
in a confidence interval $\alpha_{\min} \leq \alpha \leq \alpha_{\max}$, and
determine the associated cooling domains. This random unitary case samples axes
of rotations distributed over the $\sigma_x \cross \sigma_y$ plane. The cooling
regions  are found by integrating the polarization value of the rotated ensemble
of states,
\begin{equation}
    \pol_{\chi(\gref)} = \gamma \sqrt{1-\virpol^2} \sin(\alpha - \alpha_\rf) \sin(\chi(\gref)) + \virpol\cos(\chi(\gref)) ,
\end{equation}
over the surface defined by $\alpha$ and $\alpha_\rf$:
\begin{equation}
   \begin{split}
     \overline{\pol_{\chi(\gref)}} ={}& \frac{1}{\Delta\alpha^2} \int_{\alpha_{\min}}^{\alpha_{\max}} \int_{\alpha_{\min} - \pi/2}^{\alpha_{\max} - \pi/2} \dd{\alpha} \dd{\alpha_\rf} \pol_{\chi(\gref)} \\
     ={}& \frac{
            2 \gamma \gref (\virpol^2-1) \cos(\Delta \alpha) - 2 \gamma \gref (\virpol^2-1)+\Delta \alpha^2 \virpol^2
        }{
            \virpol^2 - (\virpol^2 - 1) \gref^2
        }
   \end{split}.
   \label{12}
\end{equation}
In Fig.~\ref{fig:alpha-interval}(a), we show the parameter space for average
achievable polarization, $ \overline{\pol_{\chi(\gref)}}$, with corresponding
cooling (blue) and heating (red) regions. We again fix the initial polarization
at $\virpol = 0.8$, as in the main text. We observe a large cooling region
(bottom part of the figure) for various values of $\Delta \alpha$ and of $\gref
/ \gamma$. Incomplete knowledge of the coherence phase  $(\Delta \alpha = 2\pi)$
does not lead to average cooling, in contrast to what happens for incomplete
knowledge of the coherence parameter $\gamma$ discussed in the main text.
Enhanced average cooling is achieved for all $\gref/\gamma$ when $\Delta \alpha
< 0.9 \pi$. We also note that choosing the coherence magnitude of the unitary
rotation $\gref$ to be smaller than its actual value $\gamma$ (that is, taking
smaller ratio $\gref/\gamma$) is beneficial for achieving improved average
cooling with large confidence interval $\Delta \alpha$, although typically one
would  cool less by doing so.

We display an example scenario of average cooling in
Fig.~\ref{fig:alpha-interval}(b) where a fully geometric representation of the
initial and final ensemble of states is shown, with $\Delta\alpha = \pi/2$ and
$\gamma = 0.5$. The applied rotation has a sharp value $\gref = \gamma$ and the
rotation axes are sampled from an ``orthogonal interval'' by setting $\alpha_\rf
= \alpha - \pi/2$ at every point of the phase confidence interval. The rotated
ensemble of states (in blue and red), with an average polarization above the
initial value of $\virpol = 0.8$, is clearly identifiable. As a guide to the
eye, the surface for absolute uncertainty is also plotted, and is seen in the
background of the surface of average cooling.

It is worthwhile mentioning that the above discussions on the achievable target polarization with incomplete information about the amount of coherence is based on a random density operator that depends on a stochastic variable ($\gamma$ or $\alpha$) specified by a probability distribution ($P(\gamma)$ or $P(\alpha)$). While a deterministic density operator  characterizes the properties of a statistical ensemble, such a random density operator can be regarded as characterizing the statistical properties of an ensemble of ensembles \cite{bre02}. This ensemble of ensembles contains more information about a quantum system than the averaged ensemble, especially concerning fluctuation properties. For instance, the probability distributions ($P(\gamma)$ or $P(\alpha)$) allow us to determine  in which cases
it is possible to guarantee enhanced cooling of the target qubit {for each
individual realizations of the experiment}, and not only on average.

\section{Coefficient of performance and cooling power}

We analyze in this section the thermodynamic performance of the generalized,
coherent three-qubit heat-bath algorithmic cooling example of the main text. The
coefficient of performance $\zeta$ is a central figure of merit for
refrigerators that is  defined as the ratio of heat extracted and work supplied
during the compression step, while the cooling power $J$ characterizes the rate
of heat extraction. For discrete cooling cycles, there are given by \cite{sol22}
\begin{equation}
    \zeta(n) = - \frac{Q(n)}{W(n)} \quad \text{ and }
    J = {|}Q(n+1) - Q(n){|}.
\end{equation}
In the presence of coherence, the definition of heat has to be extended to take
the change of coherence properly into account \cite{ber20,ssu23,rod19,ham22}.
The goal is to capture the cooling advantage provided by the manipulation of
quantum coherence.

The  heat extracted from the target qubit is concretely given by $-Q = -
\tr_1 H_1 ( \rho_1^{(n+1)} - \rho_1^{(n)} )$, whereas work is defined as the
total energy change over the target and two reset qubits during the compression
step, $W = \sum_i \tr_i H_i ( \rho_i^{(n+1)} - \rho_i^{(n)} )$,  with  $i = 1,
2, 3$. We can express these average energy changes in terms of the eigenvalues
$\omega_i$ of $H_i$ and $p_i$ of $\rho_i$, as well as in terms of the transition
amplitudes between the eigenbases of these two operators, $c_{ika}^{(n)} =
\braket{k}{n, a}_i$, where we admit possible changes in the qubit state
eigenbases, $ \ket{n, a}_i$. We accordingly obtain  for the extracted heat 
\begin{equation}
    -Q = - \tr_1\{ H_1 ( \rho_1^{(n+1)} - \rho_1^{(n)} ) \}
    = - \sum_{ka} \omega_{ik} \big[
        (\Delta p_{1a}^{(n)}) |c_{1ka}^{(n)}|^2
        + p^{(n+1)}_{1a} \Delta (|c_{1ka}^{(n)}|^2)
    \big] ,
\end{equation}
where $\Delta$ represents the cycle finite-difference of the $n$-dependent
number on which it acts, e.g.~$\Delta (|c_{ika}^{(n)}|^2) = |c_{ika}^{(n+1)}|^2
- |c_{ika}^{(n)}|^2$. This expression accounts for the possible cycle-dependence
of the qubit eigenbasis.

\begin{figure}[t]
    \centering
    \includegraphics{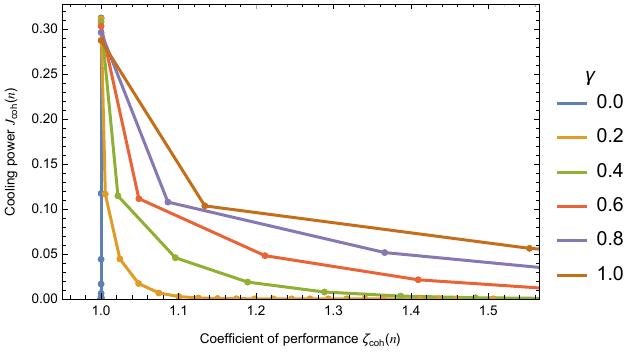}
    \caption{Cooling power $J_\textrm{coh}(n)$ against the coefficient of
    performance $\zeta_\textrm{coh}(n)$, Eq.~\eqref{17}, for various values of
    the cycle number $n$ and different amounts of coherence $\gamma$. The coefficient of
    performance is enhanced in the presence of coherence $(\gamma >0)$ compared
    to the incoherent heat-bath algorithmic cooling prootocol $(\gamma =0)$.
    {The initial target qubit polarization is $\epsilon_1(0) = 0$ and the two reset qubit
    polarizations are given by $\epsilon_2 = \epsilon_3 = 0.5$}.}
    \label{fig:cop-vs-pow}
\end{figure}

We next define the coherent-energetic quantity \cite{ber20,ssu23}
\begin{equation}
    C_i = \sum_{ka} \omega_{ik} p_{ia}^{(n+1)} \Delta |c_{ika}^{(n)}|^2 ,
\end{equation}
which is positive ($C_i > 0$) whenever the qubit $i$ acquires coherence over a
cycle. In order to include the contribution of quantum coherence to heat
exchange, we introduce  the 'coherent' heat extracted from the target qubit
\cite{ber20,ssu23}%
\begin{equation}
   - \mathcal{Q} = -Q + C_1
    = - \sum_{ka} \omega_{ik} (\Delta p_{1a}^{(n)}) |c_{1ka}^{(n)}|^2 .
\end{equation}
Since the compression step is not modified, the work  is left unchanged. We then
obtain the coefficient of performance and cooling power of the generalized
heat-bath algorithmic cooling protocol%
\begin{equation}
\label{17}
    \zeta_\textrm{coh}(n) = -\frac{\mathcal{Q}(n)}{W(n)} \quad \text{ and } \quad J_\textrm{coh}(n)  = |\mathcal{Q}(n+1) - \mathcal{Q}(n)|.
\end{equation}
Figure~\ref{fig:cop-vs-pow} shows the power $J_\textrm{coh}(n)$ against the
coefficient of performance $\zeta_\textrm{coh}(n)$ for various values of $n$ and
different amounts of coherence $\gamma$.   We note that the coefficient of performance is  enhanced while the power is reduced in the presence of coherence $(\gamma >0)$ compared to the (standard)
incoherent case $(\gamma =0)$. In particular, the coherent coefficient of
performance, $\zeta_\textrm{coh}(n)$,  can exceed the corresponding incoherent
Carnot coefficient of performance, $\zeta_\text{C}(n) =
T_\text{c}(n)/[T_\text{h}- T_\text{c}(n)]$, with  the cold temperature, $
T_\text{c}(n) = \ln[ ( 1 + \epsilon_1(n) ) / ( 1 - \epsilon_1(n) ) ]$
determined by the polarization of the target qubit, and  the hot temperature
$T_\text{h}$ set by the heat bath. This behavior is similar to that observed for
a refrigerator in Ref.~\cite{ham22}.

\vspace{1cm}

\section{Comparison with other resources}

\begin{figure}[t!]
    \centering
    \includegraphics{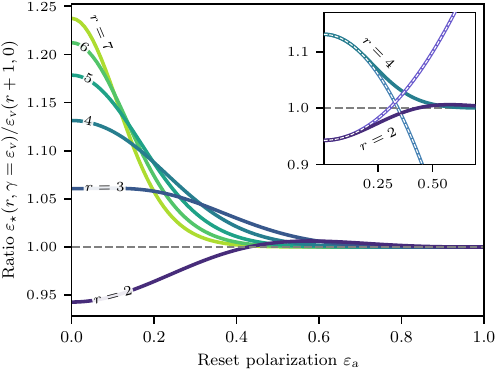}
    \caption{{Quantum coherence versus one additional reset qubit. Ratio
    $\epsilon_\star(r,\gamma = \epsilon_\vir)  / \epsilon_\vir(r+1,\gamma=0)$ of
    the maximum polarization attainable with $r$ reset qubits by adding
    coherence, $\gamma = \epsilon_\vir$, and the maximum polarization obtainable
    by adding one more reset qubit without coherence ($\gamma=0$), as a function
    of the reset polarization $\epsilon_\anc$, for various values of $r$. The
    coherent cooling scheme is better for moderate polarizations $\epsilon_\anc$
    for $r=2$ and for small polarizations $\epsilon_\anc$  for $r \geq 3$. For
    small reset polarizations, the ratio depends quadratically on
    $\epsilon_\anc$ (inset).}}
    \label{fig:reset-comparison}
\end{figure}
Another way to increase the target polarization in incoherent heat-bath
 algorithmic cooling is to augment the number of reset qubits \cite{rod16}. The
 asymptotic polarization grows exponentially with the number $r$ of reset qubits
 as $\epsilon_\infty(r, \epsilon_\anc) = [(1 + \epsilon_\anc)^r - (1 -
 \epsilon_\anc)^r]/[(1 + \epsilon_\anc)^r + (1 - \epsilon_\anc)^r]$, when all
 the reset qubits  have the same polarization $\epsilon_\anc$ \cite{rod16}. It
 is hence instructive to compare the cooling improvement achieved by the two,
 coherent and incoherent, methods. To that end, we consider a cooling
 algorithm with $r$ reset qubits \cite{com3} and define the ratio
 $\epsilon_\star(r,\gamma = \epsilon_\vir) / \epsilon_\vir(r+1,\gamma=0)$ of the
 maximum polarization attainable by adding the largest amount of coherence,
 $\gamma= \epsilon_\vir$, to the corresponding virtual qubit and the maximum
 polarization obtainable by adding one more reset qubit in the absence of
 coherence ($\gamma=0$). Figure~\ref{fig:reset-comparison} displays this ratio
 as a function of the reset polarization $\epsilon_\anc$, for different values
 of $r$. We observe that, for two reset qubits ($r=2$), adding coherence leads
 to a larger polarization than adding one reset qubit for moderate reset
 polarizations ($0.45<\epsilon_\anc<0.9$), while both schemes are equivalent for
 high reset polarizations ($0.9<\epsilon_\anc$). On the other hand, for $r\geq
 3$, coherence appears to be always beneficial for smaller reset polarization,
 with an enhancement of more than $10\%$ for $r \geq 4$.


\begin{thebibliography}{99}
\bibitem{com} A virtual quantum subsystem $\tilde S$ of a larger system $E$ is associated with a tensor factor ${\cal H}_{\tilde S}$ of a subspace of  ${\cal H}_{E}$ such that ${\cal H}_{E} = ({\cal H}_{\tilde S}\otimes {\cal H}_{F}) \oplus {\cal H}_{R}$, for some factor ${\cal H}_{F}$ and a remainder space ${\cal H}_{R}$, see, for example, Ref.~\cite{tic14}.
\bibitem{com1} We emphasize that the final unitary rotation applies to the qubit system and not to the virtual qubit. This is experimentally advantageous since virtual qubits are often encoded in complex multipartite systems.
\bibitem{com4} This amounts to an inexact diagonalization of the virtual state, in contrast to the exact diagonalization that lead to the optimal bound \eqref{2}.
\bibitem{com2} The worst case overall is for the pair of values
$(\gamma = 0, \gref = 1)$, where the state is heated to $\pol_{\chi(1)} =
\virpol^2 < \virpol$.
\bibitem{com3} The virtual qubit of the incoherent heat-bath algorithmic cooling protocol with $r$ reset qubits of Ref.~\cite{rod16} is given by the two states $\ket{00\dots 0}$ ($r$ times $0$) and $\ket{11\dots 1}$ ($r$ times $1$) of the multi reset qubit system.
\end{thebibliography}
\end{document}